\documentclass[%
 reprint,
 superscriptaddress,
 amsmath,amssymb,
 aps,
 prl
]{revtex4-2}

\usepackage{graphicx}
\usepackage{dcolumn}
\usepackage{bm}
\usepackage{hyperref}

\usepackage{amssymb}
\usepackage{amsmath,bm}
\usepackage{float}
\usepackage[usenames,dvipsnames]{color}
\usepackage[export]{adjustbox}
\usepackage{gensymb}

\usepackage{soul,xcolor}

\usepackage{multirow}
\graphicspath{{figures/}} 
\usepackage{tikz}
\usepackage[normalem]{ulem}

\newcommand{\ket}[1]{\left|  #1 \right\rangle}

\newcommand{\aver}[1]{\ensuremath{\langle {#1} \rangle}}

\newcommand{\spinup}[0]{\ensuremath{\ket{\uparrow}}}
\newcommand{\spindown}[0]{\ensuremath{\ket{\downarrow}}}

\usepackage{graphicx}
\usepackage{dcolumn}
\usepackage{bm}
\usepackage{hyperref}

\usepackage{amssymb}
\usepackage{amsmath,bm}

\begin{document}

\title{
Fast Preparation and Detection of a Rydberg Qubit using Atomic Ensembles}

\author{Wenchao Xu}
\thanks{These authors contributed equally}
 \affiliation{Department of Physics and Research Laboratory of Electronics, Massachusetts Institute of Technology, Cambridge, Massachusetts 02139, USA}
 
\author{Aditya V. Venkatramani}
\thanks{These authors contributed equally}
\affiliation{Department of Physics, Harvard University, Cambridge, Massachusetts 02138, USA}
\affiliation{Department of Physics and Research Laboratory of Electronics, Massachusetts Institute of Technology, Cambridge, Massachusetts 02139, USA}
 
\author{Sergio H. Cant\'{u}}
\thanks{These authors contributed equally}
\affiliation{Department of Physics and Research Laboratory of Electronics, Massachusetts Institute of Technology, Cambridge, Massachusetts 02139, USA}
 
\author{Tamara \v{S}umarac}
\affiliation{Department of Physics, Harvard University, Cambridge, Massachusetts 02138, USA}
\affiliation{Department of Physics and Research Laboratory of Electronics, Massachusetts Institute of Technology, Cambridge, Massachusetts 02139, USA}

\author{Valentin Kl\"usener}
\affiliation{University of Erlangen-Nuremberg, Germany}
\affiliation{Department of Physics and Research Laboratory of Electronics, Massachusetts Institute of Technology, Cambridge, Massachusetts 02139, USA}

\author{Mikhail D. Lukin}
\affiliation{Department of Physics, Harvard University, Cambridge, Massachusetts 02138, USA}

\author{Vladan Vuleti\'{c}}
\affiliation{Department of Physics and Research Laboratory of Electronics, Massachusetts Institute of Technology, Cambridge, Massachusetts 02139, USA}

\begin{abstract}
We demonstrate a new approach for fast preparation, manipulation, and collective readout of an atomic Rydberg-state qubit. By making use of Rydberg blockade inside a small atomic ensemble, we prepare a single qubit within 3~$\mu$s with a success probability of $F_p=0.93 \pm 0.02$, rotate it, and read out its state in $6$ $\mu s$ with a single-shot fidelity of $F_d=0.92 \pm 0.04$. The ensemble-assisted detection is $10^3$ times faster than imaging of a single atom with the same optical resolution, and enables fast repeated non-destructive measurement. We observe qubit coherence times of 15~$\mu$s, much longer than the $\pi$ rotation time of 90~ns. Potential applications ranging from faster quantum information processing in atom arrays to efficient implementation of quantum error correction are discussed.

\end{abstract}

\maketitle

Fast and reliable state initialization and readout of qubits are essential requirements for implementing scalable quantum information systems. Recently, individually-controlled highly excited Rydberg atoms have emerged as a promising platform for quantum simulation and computation \cite{lukin2001dipole,browaeys2020many,saffman2016quantum}. These systems are enabled by the strong coherent interaction between Rydberg atoms at distances exceeding several micrometers. 
In combination with the demonstrated ability to deterministically assemble large arrays of individual atoms  \cite{endres2016atom,barredo2016atom,barredo2018synthetic,kumar2018sorting, de2019defect,brown2019gray,kim2016situ}, Rydberg-atom arrays have been used to simulate quantum spin models \cite{browaeys2020many} with more than 250 qubits \cite{bernien2017probing,keesling2019quantum,Ebadi2020,Scholl2020} to perform multiple-qubit gate operations \cite{isenhower2010demonstration,graham2019rydberg,levine2019parallel,jaksch2000fast,levine2018high,madjarov2020high}, or to create large maximally entangled states \cite{omran2019generation}. While these quantum simulation and computation systems can operate on microsecond time scales, they could benefit substantially from faster qubit preparation and detection, as both the array preparation process and the optical state readout in most systems require several to many milliseconds~\cite{endres2016atom,barredo2016atom,barredo2018synthetic}. Moreover, fast and high-fidelity single-shot qubit readout without atom loss could enable a new generation of experiments with error mitigation, such as quantum error correction and fault tolerant quantum processing~\cite{devitt2013quantum}.

Prior approaches for individual Rydberg-qubit detection include state-dependent ionization and detection of the ions, a relatively fast ($\tau \sim 0.1$~ms) process that has only moderate fidelity \cite{gallagher2005rydberg}, and the state-dependent removal of atoms followed by relatively slow ($\tau \sim 10$~ms) fluorescence imaging of the remaining atoms \cite{gaetan2009observation,bernien2017probing,madjarov2020high,Fuhrmanek2010} with relatively high fidelities of $F \gtrsim 0.95$. Fast high-intensity fluorescence detection within 20~$\mu$s with single-atom resolution has been achieved in Ref. \cite{Bergschneider2018}; however, this method as demonstrated is not compatible with atomic arrays, as it does not have the necessary spatial resolution and also requires a large magnetic field. Both ion detection and fluorescence imaging are destructive readout processes, and require a new atomic array to be prepared subsequently, further limiting the cycle time of the quantum processor.

Our detection scheme is based on the proposal by G\"unter {\it et al.} \cite{gunter2012interaction} to use Rydberg interactions in combination with electromagnetically induced transparency (EIT) \cite{fleischhauer2005electromagnetically,mohapatra2007coherent} for collectively enhanced imaging. This method has previously been used to observe Rydberg dynamics \cite{gunter2013observing} without, however, experimentally achieving single-atom resolution. A similar approach has also been used to demonstrate a single-photon switch and a single-photon transistor using Rydberg interactions~\cite{baur2014single,gorniaczyk2014single,tiarks2014single}.

\begin{figure}[htbp]
\centering
\includegraphics[width=0.48\textwidth,trim= 150 215 180 
85,clip]{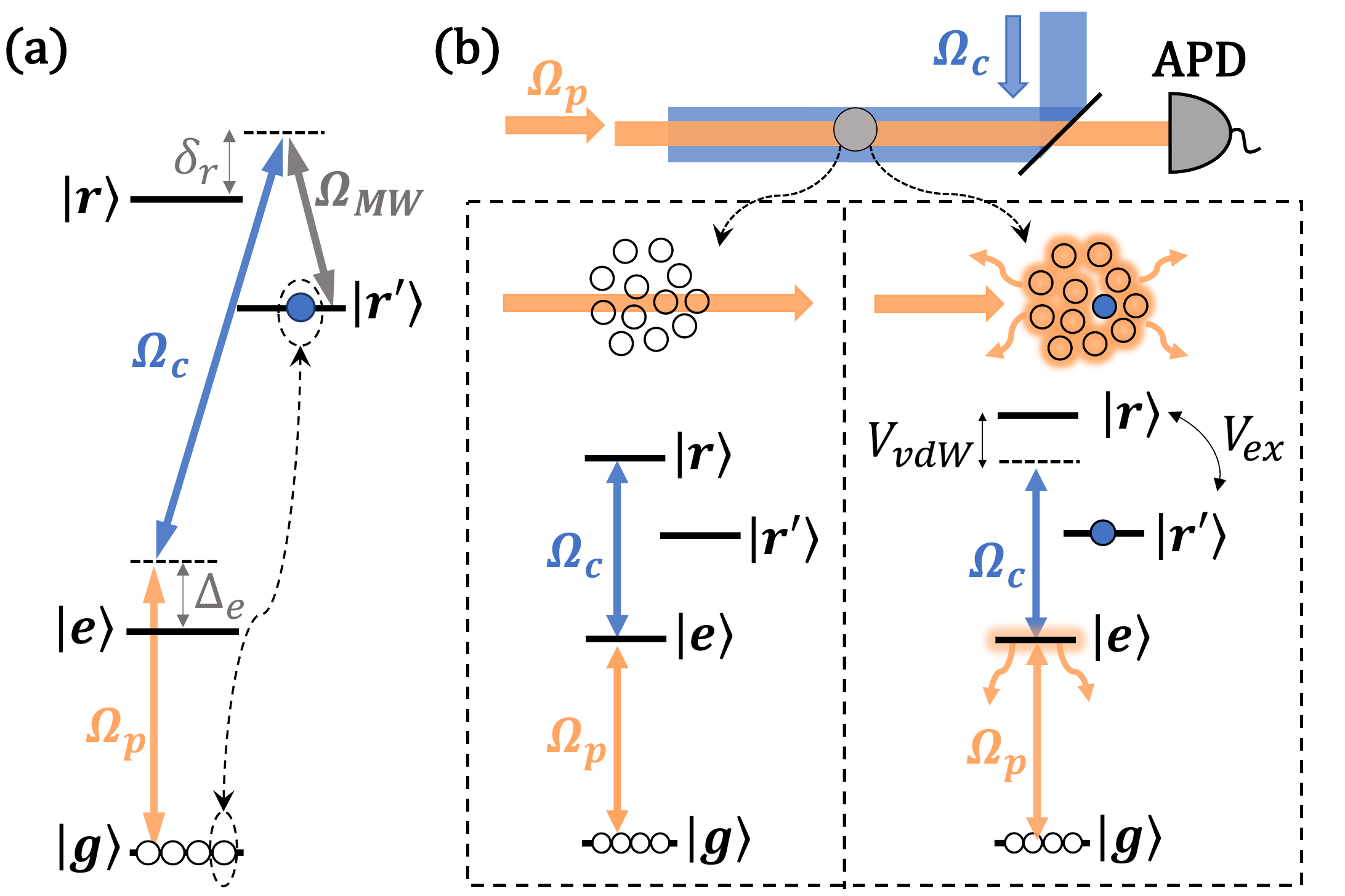}
\caption{\label{fig:Setup}
Fast collective detector of a single Rydberg atom.
a. State initialization. An atom is prepared in the Rydberg state $\ket{r'}$ through a three-photon process involving the preparation beam ($\Omega_p$, orange), the control beam ($\Omega_c$, blue), and a microwave field ($\Omega_{MW}$, grey). The detunings from the two intermediate states are $\Delta_e = \delta_r = 2\pi \times 100$~MHz. The preparation of a single atom in $\ket{r'}$ is ensured by the strong interaction between two atoms in $\ket{r'}$  \cite{lukin2001dipole}. 
b. A probe field (orange, waist size  $w_p=4.5$~$\mu$m) in combination with the control field ($w_c=12.5$~$\mu$m) couples atoms to the Rydberg state $\ket{r}$. Under conditions of EIT ($\Delta_e=\delta_r=0$), high transmission through the atomic medium results in a large number of detected photons (left). On the other hand, if the Rydberg state $\ket{r'}$ is populated by an atom (right), then the strong interaction between $\ket{r}$ and $\ket{r'}$ removes the EIT condition, resulting in a significant reduction of transmitted photon number due to absorption by the ensemble. The interaction $V_{rr'}$ contains both dipolar-exchange ($V_{ex}$) and van-der-Waals components ($V_{vdW}$) (see SM). }
\end{figure}

In this Letter, we demonstrate high-fidelity preparation, manipulation and detection of a single-Rydberg-atom qubit (and not a collective state as in Ref. \cite{Ebert2015}) inside an atomic ensemble on the microsecond time scale. Starting with $N \sim 400$ trapped ultracold $^{87}$Rb atoms, we prepare a qubit between the Rydberg states $\spinup \equiv \ket{r'} = \ket{91P_{3/2},m_j=3/2}$ and $\spindown \equiv \ket{r} = \ket{92S_{1/2}, m_j=1/2}$, perform qubit rotations with a loss of contrast $\delta C \leq 2 \times 10^{-3}$ per $2 \pi$ pulse, and read out the state optically. Harnessing the collective effect of Rydberg blockade \cite{jaksch2000fast}, the state preparation and detection are performed in $T_p=3~\mu$s and $T_d=6~\mu$s with fidelities of $F_p=0.93 \pm 0.02$ and $F_d=0.92 \pm 0.04$, respectively. The qubit coherence time of $\tau_c = (15 \pm 5)$~$\mu$s, measured with a Ramsey sequence, is much longer than the $\pi$ rotation time of 90~ns.

Our approach harnesses collective phenomena for speeding up both state preparation and detection. The preparation is accomplished by applying laser and microwave radiation
to an ensemble of $N$ atoms, such that any atom can be excited to the Rydberg state, yielding $N$ times faster excitation of the first atom to the Rydberg state than for a single atom, while the preparation of a single excitation is ensured by the Rydberg blockade mechanism \cite{urban2009observation,bernien2017probing}. Similarly, the signal-to-noise ratio in optical detection is collectively enhanced by a factor $\sim N$: Depending on the state of the single-atom Rydberg qubit, the absorption of probe light by all of the $N$ atoms in the ensemble is simultaneously switched on or off \cite{gunter2012interaction}. 

Our experimental setup is illustrated in Fig. \ref{fig:Setup}a. A small ensemble with root-mean-square (rms) size of $\sqrt{\aver{r^2}}  \approx 6$~$\mu$m  containing typically $N \sim 400$  laser-cooled $^{87}$Rb atoms is prepared inside a two-beam optical dipole trap with waist sizes $w_1=10~\mu$m and $w_2=20 ~\mu$m (see Supplemental Material (SM) \cite{SM} for details). The trapped atoms are optically pumped into the hyperfine and magnetic sublevel $\ket{g} \equiv \ket{5S_{1/2}, F=2, m_F=2}$ that is coupled via a two-photon process involving the transitions $\ket{g} \leftrightarrow \ket{e} \equiv \ket{5P_{3/2}, F=3, m_F=3}$ (probe beam $\Omega_p$) and $\ket{e} \leftrightarrow \ket{r}$ (control beam $\Omega_c$) to the Rydberg state $\ket{r} \equiv \ket{92S_{1/2}, m_j=1/2}$.  

To prepare a single atom in the Rydberg state $\ket{r'}$ inside the ensemble, the probe laser and microwave field are detuned by $\Delta_e/(2\pi)= \delta_r/(2\pi)=100$~MHz from their respective transitions, and in combination with the control field drive a three-photon transition $\ket{g} \leftrightarrow \ket{e} \leftrightarrow \ket{r} \leftrightarrow \ket{r'}$ (see Fig. \ref{fig:Setup}a). By changing the powers of the two optical fields within $\sim$3$~\mu$s, while keeping the microwave field constant, a process similar to stimulated Raman adiabatic passage (STIRAP) is realized (see SM \cite{SM} for details). This process is chosen over direct excitation because it 
is less sensitive to laser noise and atom number fluctuations. The observed linewidth $\Gamma_3/(2\pi)=0.6$~MHz of the three-photon transition is much smaller than the energy shift $|\Delta E|/h \gtrsim 10$~MHz at the rms distance $d_0 \equiv \sqrt{2\aver{r^2}} \approx 8$~$\mu$m between two atoms in their $\ket{r'}$ state in the ensemble (averaged over angles, see SM \cite{SM}). This ensures that excitations of two or more atoms to the Rydberg state are suppressed \cite{urban2009observation,bernien2017probing}. While the $\ket{r'}$ state has vanishing interactions along certain angular directions (see SM \cite{SM}), the admixture of the spherically symmetric $\ket{r}$ state with the microwave field during preparation may help increase the preparation fidelity. 

\begin{figure*}[htbp]
\centering
\includegraphics[width=0.99\textwidth,trim= 0 170 30 
150,clip]{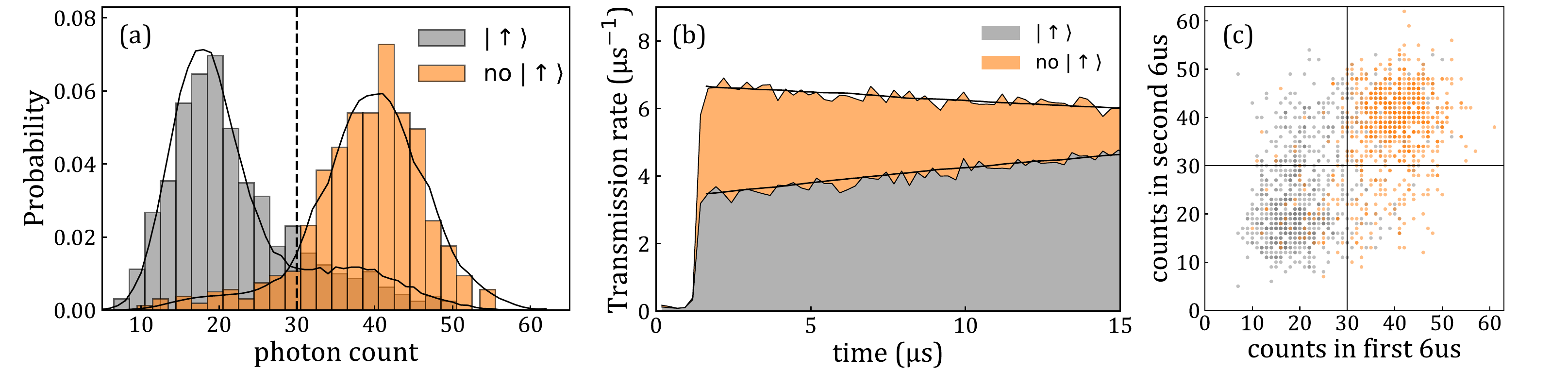}
\caption{\label{fig:Detector}
a. Histogram of the transmitted probe photon number for state detection performed in 6~$\mu$s.
Grey and orange histograms correspond to the presence and absence of an atom in Rydberg state $\ket{r'} \equiv \spinup$, respectively. The solid lines in a, b indicate a theoretical model that for the presence (absence) of an atom in $\spinup$ assumes random sudden loss of the Rydberg atom in $\spinup$ (sudden decay of the slow-light polariton into a Rydberg state) at a rate $0.035$ $\mu s^{-1}$ ($0.015$ $\mu s^{-1}$). The dashed line indicates the detection threshold that gives us the highest fidelity of differentiating two distributions. 
The control Rabi frequency is $\Omega_c/(2\pi) = 25$~MHz, and the probabilities for collecting and detecting a transmitted probe photon are 0.90 and 0.47, respectively.  (b) Time-resolved photon count rate during detection. (c) Correlation plot of number of detected-photon counts in two consecutive 6$\mu$s measurements in the same run of the experiment. Gray (orange) points represent transmission data when we prepare (do not prepare) the $\spinup$ state. Vertical and horizontal lines represent threshold counts for state discrimination.
}
\end{figure*}

Before discussing how we experimentally verify that only a single Rydberg atom has been prepared, we first describe the detection process. In the following, we associate the Rydberg state $\ket{r'}$ with  an effective  $\spinup$ state. When the two-photon transition is resonant with the intermediate state ($\Delta_e=0$, see Fig. \ref{fig:Setup}a), the transmitted probe light serves for Rydberg state detection \cite{gunter2012interaction} under conditions of EIT ~\cite{fleischhauer2005electromagnetically,mohapatra2007coherent}. If an atom in $\spinup$ is present, the excitation of a Rydberg polariton to the state $\ket{r}$ at a distance $R$ requires an additional interaction energy that in the presence of both van der Waals and exchange interactions is approximately given by $V_{rr'}=C_6/R^6 \pm C_3/R^3$, where $C_{6}/h=6310~\text{GHz~} \mu\text{m}^{6}$ and $C_{3}/h=23.6~\text{GHz~} \mu\text{m}^{3}$ (see SM \cite{SM}). This interaction energy shifts the EIT resonance and results in a lower transmission of the probe beam for the state $\ket{\uparrow}$.

Fig. \ref{fig:Detector}a shows the observed photon count histograms of the transmitted light in a 6-$\mu$s detection window with and without an atom in $\spinup$. Even in such a short time, the two photon count distributions can be clearly distinguished. The time-resolved average count rate (Fig. \ref{fig:Detector}b) reveals that the transmission $T_{\spinup}$ for $\spinup$ increases with time, whereas the high transmission without an atom in $\spinup$ is almost constant, and decreases only slowly. The latter may be explained by a decay of the slow-light $\ket{r}$-polaritons to other Rydberg states, producing randomly a stationary atom in some Rydberg state, that then blocks the EIT transmission. The increase in the average transmission $T_{\spinup}$, on the other hand, can be explained by a light-induced loss process of the Rydberg atom in $\spinup$ during detection, which leads to a sudden increase in transmission at a random time. The observed signal reduction is primarily associated with the control light, and is too fast to be explained by photoionization \cite{saffman2005analysis, anderson2013ionization} or the repulsive ac-Stark shift (see SM \cite{SM} for details). We hypothesize that a small residual electric field of a few 10~mV/cm mixes a $\ket{92S}$ component into the $\ket{91P}$ state, so that the control light can couple the $\spinup$ state, which contains a small component of $\ket{92S}$ in it, to the rapidly decaying $\ket{e}$ state, causing a sudden loss of the atom in $\spinup$.


Modelling the system as a removal of blockade at a random time yields excellent agreement with the photon count histograms observed at different detection times (see SM \cite{SM}).  From this we infer a preparation fidelity for the state $\spinup$ (i.e., an atom in $\ket{r'}$) of $F_p=0.93 \pm 0.02$ (see SM \cite{SM} for details). The detection fidelity (probability of correctly identifying the underlying state $\spinup$) after removing the state preparation error is then $F_d=0.92 \pm 0.04$. 

Fig. \ref{fig:Detector}c demonstrates that we can perform repeated ('non-destructive') measurements on the system, where a second 6-$\mu$s measurement yields good agreement with the first measurement: The average conditional probability for the second measurement to have the same outcome as the first measurement is $p=0.79 \pm 0.03$ (see SM \cite{SM} for details). The detection system can also be viewed as a single-atom transistor for light. We then achieve a gain of $G=17 \pm 1$ in 6~$\mu$s, somewhat larger than the gain of $G=10$ in 30~$\mu$s achieved in the Rydberg system of Ref. \cite{gorniaczyk2014single}.

We implement a qubit in our system by defining the state with a single atom in $\ket{r}$ as the $\spindown$ state. Coherent rotations in the $\{\spinup,\spindown \}$ manifold are induced by applying the microwave field. After a qubit rotation, we detect the resulting state by turning on the coupling light slightly (1~$\mu$s) earlier than the probe light, such that the state $\ket{r}$ is quickly de-excited by the strong coupling laser to the unstable state $\ket{e}$, which decays by photon emission in 30~ns (see Fig. \ref{fig:Setup}). Thus, as far as the detection process is concerned, the state $\spindown$ (atom in $\ket{r}$) is equivalent to having no Rydberg excitation at all, while the state $\spinup$ (atom in $\ket{r'}$) remains unaffected by the detection light, and leads to Rydberg blockade of the probe transmission. If the photon count is above or below a chosen detection threshold (see Fig. \ref{fig:Detector}a), we identify the qubit state as $\spindown$ or $\spinup$, respectively.

\begin{figure}[htbp]
\centering
\includegraphics[width=0.48\textwidth,trim= 100 110 180 
70,clip]{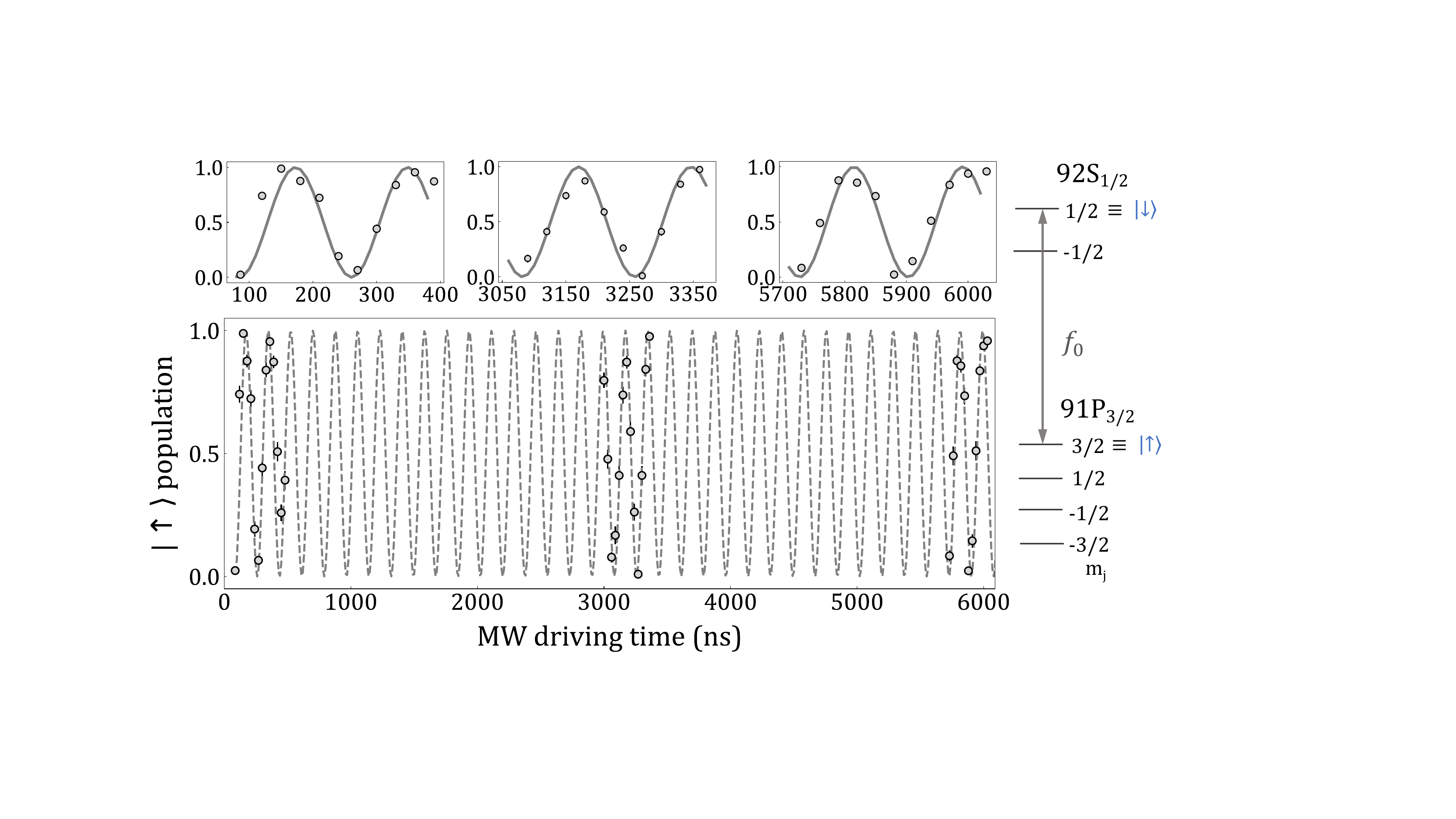}
\caption{\label{fig:Rabi}
A microwave field at a frequency $f_{0}=4814.2$~MHz is applied to drive Rabi oscillations between $\spinup$ and $\spindown$ at an oscillation frequency $\Omega/(2\pi) = 5.3$~MHz. Each point is an average of $\sim 150$ repetitions. The error bars are the standard deviation of the mean. The fitted contrast loss per $2\pi$ pulse is $\delta C < 2 \times 10^{-3}$. The relevant energy level diagrams are shown on the right. 
}
\end{figure}

Fig. \ref{fig:Rabi} shows Rabi oscillations with the full sequence of state preparation, qubit rotation, and detection.
The trap light is turned off during the sequence to avoid light shifts of the states. We use two microwave antennas with adjusted relative phase and amplitude to suppress the $\pi$ polarization component of the microwave field that can couple atoms in $\ket{\downarrow}$ to the magnetic sublevel $m_j=1/2$ in the $91P_{3/2}$ manifold, offset by $17$~MHz in an applied magnetic field of $9$~G. The remaining coupling to other magnetic sublevels limits the maximum Rabi frequency on the $\spinup \leftrightarrow \spindown$ transition to $\lesssim 5$~MHz. The Rabi oscillations show no observable damping on the 6~$\mu$s timescale, corresponding to a contrast loss per $2\pi$ pulse of $\delta C \leq 2 \times 10^{-3}$.

The observed contrast of the Rabi oscillations can be used to determine the probability that two excitations in $\ket{r'}$ were simultaneously created in the ensemble. Due to the large interaction energy between two atoms in $\ket{r'}$ and $\ket{r}$, the Rabi oscillations with two excitations would very quickly wash out on a time scale $h/V_{rr'}(d_0) \sim 40$~ns. From the observed contrast of the Rabi oscillation we conclude that the probability for preparing two excitations is below $ 1\%$.

\begin{figure}[htbp]
\centering
\includegraphics[width=0.5\textwidth,trim= 50 0 100 20,clip]{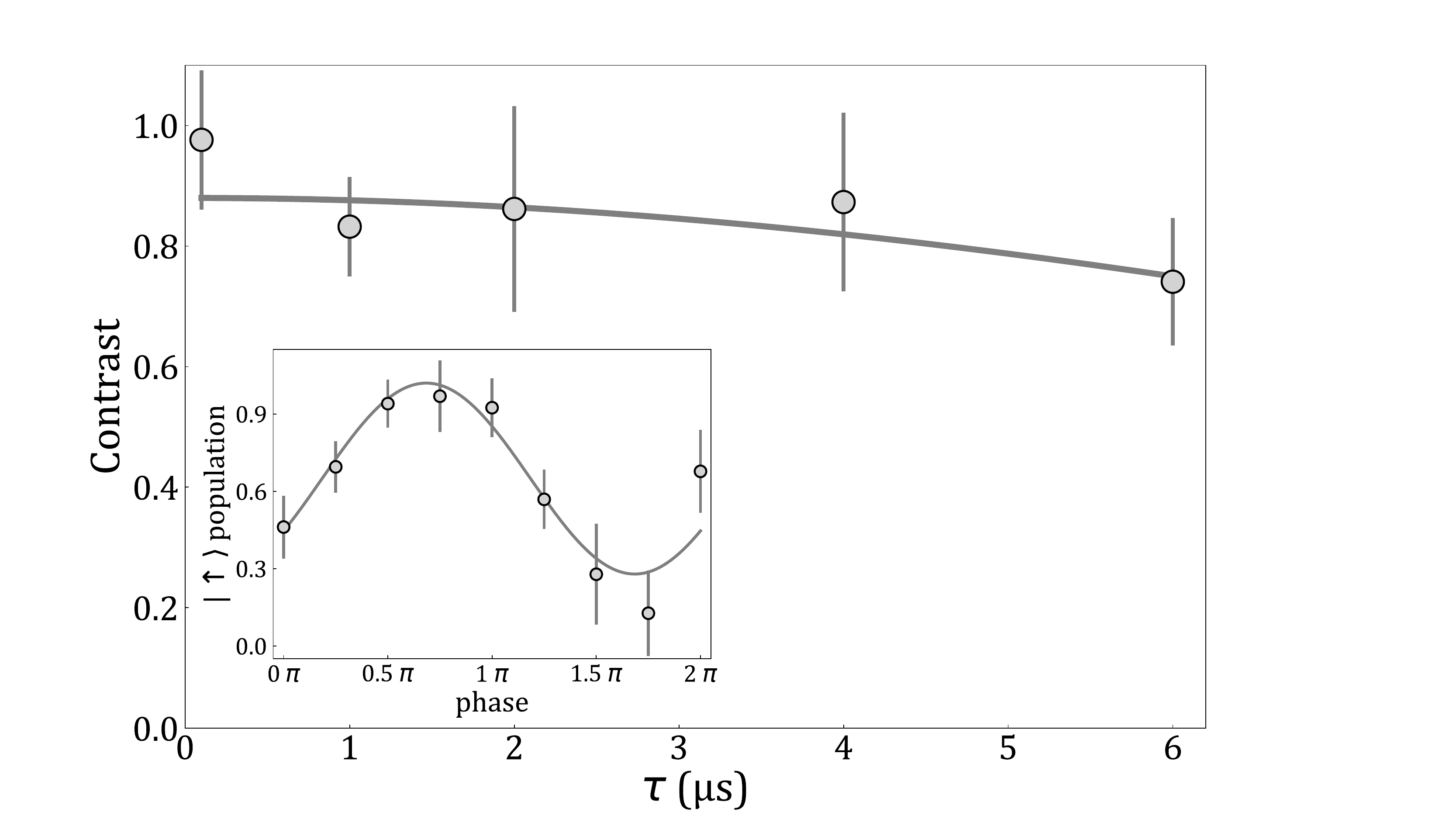}
\caption{\label{fig:Ramsey}
Ramsey measurements consisting of two $\pi/2$ pulses separated by a time $\tau$. The phase of the second $\pi/2$ pulse is scanned to obtain a Ramsey fringe. The contrast of the Ramsey fringe is plotted as a function of $\tau$. The solid curve is a fit to a Gaussian decay function $A e^{-(\tau/T_2^{*})^{2}}$, yielding a dephasing time $T_2^* = (15 \pm 5)$~$\mu$s and $A=0.88 \pm 0.04$. Inset: Ramsey fringe at $\tau=1000$~ns. 
}
\end{figure}

We use a Ramsey measurement to characterize the coherence time of the Rydberg qubit embedded inside the atomic cloud. Two $\pi/2$ pulses are applied with a temporal separation $\tau$ between them, and their relative phase is scanned to obtain a Ramsey fringe at given $\tau$. Fig.~\ref{fig:Ramsey} displays the contrast of the Ramsey fringes as a function of Ramsey time $\tau$. By fitting the contrast to a Gaussian decay function, we obtain the $e^{-2}$ dephasing time as $(15 \pm 5)$~$\mu$s. Possible dephasing mechanisms include electric-field fluctuations acting on the highly polarizable Rydberg states,
magnetic field fluctuations, and interactions between the Rydberg atom and the surrounding ground state atoms \cite{PhysRevX.6.031020}. We also note that neither the Rabi flopping nor the Ramsey measurement depend on whether we have encoded the qubit in a single atom or collectively in a W-state \cite{Ebert2015}. However, previous measurements involving storage and retrieval of photons indicate that the collective state decoheres during preparation, such that the qubit is ultimately encoded in a single atom \cite{peyronel2012quantum}.

In summary, by harnessing collective effects in a small atomic ensemble, we have demonstrated a method for the rapid preparation and detection of a Rydberg qubit. The preparation and detection fidelities demonstrated in this work can likely be further increased in the future. The preparation fidelity for a single excitation can be improved by modifying the preparation sequence (see SM \cite{SM}) or using smaller ensemble size, since such ensembles would provide even higher energy cost for multiple excitation. The size of the ensemble cannot, however, be made arbitrarily small, since at higher atomic densities, necessary to maintain the same optical depth $OD \sim 1$, Rydberg molecule formation \cite{bendkowsky2009observation} could lead to loss. Given our current average atomic density of $\aver{n}=2 \times 10^{11}$~cm$^{-3}$, reducing the ensemble size by a factor of 2 should be possible, which would likely reduce the preparation error by more than an order of magnitude.

Our demonstrated detection fidelity, on the other hand, is limited by the loss of the Rydberg atom prepared in the $\ket{r’}$ state. This loss is mainly caused by the control light in the detection stage, and thus could be mitigated by using two ensembles, one for hosting the qubit and the other for detection, located within a blockade radius from each other. Assuming a measurement time of 10~$\mu$s, this configuration could allow for a non-destructive, fast qubit readout with detection fidelity over $99\%$, a crucial tool necessary for implementing quantum error correction \cite{devitt2013quantum}. In addition, such a readout can also enable studies of quantum feedback~\cite{sayrin2011real}, quantum Zeno effect~\cite{barontini2015deterministic}, quantum jumps~\cite{minev2019catch}, and can act as a fast probe of Rydberg super-atom dynamics \cite{PhysRevX.5.031015}.

The detection scheme can be readily implemented in different Rydberg platforms \cite{keesling2019quantum,browaeys2020many,barredo2018synthetic}, where a single atom would be replaced by a small ensemble, as demonstrated in Ref.~\cite{wang2019preparation}. Alternatively, one can place a small ensemble within the Rydberg blockade radius of each single atom for fast detection, or even within the blockade radius of several single atoms for fast parity measurements, and possibly even error correction. To suppress diffusion of the Rydberg atom between different ensembles due to the exchange interaction during detection, it may be necessary to adjust the lattice constant of the array or the principal quantum numbers of the Rydberg states used.

We thank Annika Tebben, Rivka Bekenstein, and Florentin Reiter for discussions. This work has been supported in part by NSF, the NSF-funded Center for Ultracold Atoms, ARO, ARL, AFOSR, DARPA, DOE and Boeing.

\bibliography{reference}
\end{document}



\newpage

\section{supplementary material}
\subsection{Experimental setup}
$^{87}$Rb atoms are collected in a three-dimensional magneto-optical trap (MOT) and loaded into a crossed optical dipole trap created with two orthogonal far-detuned laser beams at wavelengths 852~nm (propagating along $z$) and 1064~nm (propagating along $y$) with waist sizes 10~$\mu$m and 20~$\mu$m, respectively, providing individual trap depths of typically $U_{852}/h=2$~MHz and $U_{1064}/h=20$~MHz. The trap vibration frequencies are $\omega_x/(2\pi)=5.7$~kHz, $\omega_y/(2\pi)=3.0$~kHz and $\omega_z/(2\pi)=4.8$~kHz. The probe propagates in the $xy$ plane at an angle of $16\degree$ to the $y$ axis. The cloud is cooled to $80$ $\mu K$ using polarization gradient cooling, resulting in root-mean-squared (RMS) sizes of $x_0=2.4$~$\mu$m, $y_0=4.6$~$\mu$m, $z_0=2.9$~$\mu$m for the ensemble. The rms distance between two atoms in the ensemble is then $d_0=\sqrt{2(x_0^2+y_0^2+z_0^2)}=8.4$~$\mu$m. The total absorption on the $\ket{5S_{1/2}, F=2, m_F=2} \rightarrow \ket{5P_{3/2}, F=3, m_F=3}$ transition corresponds to $N_0=440$ atoms if we assume that the atomic cross section is reduced by various broadening mechanisms by a factor of 2 from its maximum value $\sigma_0=3\lambda_0^2/(2\pi)$, where $\lambda_0=780$~nm is the wavelength of the probe transition. We apply a magnetic field of $9$ G along the direction of propagation of the probe beam to define the quantization axis and split the magnetic sublevels of the Rydberg states. The optical dipole traps are turned off during state preparation, rotation, and detection to prevent broadening of the transition due to inhomogeneous light shifts of the ground and Rydberg states. 


\subsection{Preparation of an atom in the Rydberg state}

\begin{figure}[htbp]
\includegraphics[width=\textwidth]{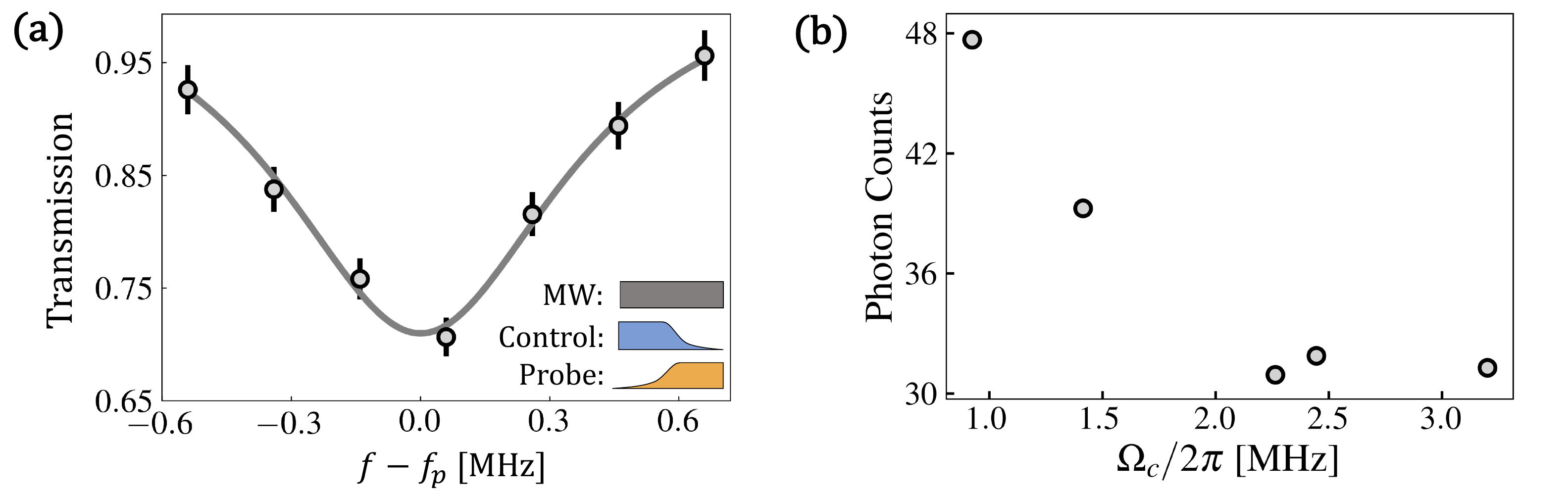}
\caption{\label{fig:RydOpt}
\textbf{Optimization of Rydberg Preparation :} (a) Spectrum of the 3-photon resonance absorption to the $\ket{\uparrow}$ state vs. microwave frequency relative to $f_p = 4916.1$~MHz. Atoms are prepared in the Rydberg state $\ket{\uparrow}$ by means of a three-photon STIRAP-like process.  (b) Photon counts integrated over a $10~\mu$s window in detection stage as a function of $\Omega_{c}$, the coupling of the $\ket{e}\rightarrow\ket{\downarrow}$ transition during the preparation stage. For small $\Omega_c$, the low coupling rate leads to a small population in the Rydberg state. As $\Omega_c$ increases, the average preparation to $\ket{\uparrow}$ increases as well. Once the whole ensemble is blockaded photon counts reduce by $34\%$. To avoid multiple excitations, We choose to work with $\Omega_c = 2\pi \times 2.4$~MHz, the value of which corresponds to the beginning of the saturated transmission level. }
\end{figure}

The preparation of the atom in the Rydberg state $\ket{\uparrow} \equiv \ket{91P_{3/2},m_j=3/2}$ is performed with a three-photon STIRAP-like process as described in the main text. The detuned microwave field mixes a small component of $\ket{r}$ into $\ket{r'}$, and the ramping of the two optical fields allows an adiabatic passage towards the target dressed state. To optimize the state initialization stage, we fix the frequencies of the probe and control laser beams, and scan the microwave frequency. Fig. \ref{fig:RydOpt}a displays the transmission of probe light during the detection stage as a function of the microwave frequency used for state preparation. We choose $f_p = 4916.1$~MHz that maximizes the preparation probability. 

The Rydberg blockade effect is essential for limiting the Rydberg excitation to one atom inside the ensemble. Since the blockade radius is proportional to $\Omega_c^{-1/3}$, the probability to have more than one excitation inside the ensemble will increase if the control Rabi frequency $\Omega_c$ is too high. We optimize $\Omega_c$ by monitoring the transmission of the probe light during the detection stage (Fig. \ref{fig:RydOpt}b), and choose the value $\Omega_c = 2\pi \times 2.4$~MHz that corresponds to the onset of saturation. In the future, we can further improve the preparation fidelity by a modified preparation scheme, where we ramp down the microwave power at the end of the preparation process, to avoid a small possibility of projecting the state onto $\ket{r}$ due to the sudden turn-off.

\subsection{Loss of the Rydberg atom and creation of Rydberg impurity during detection}

As shown in Fig. 2(b) in the main text, the rising slope of the transmission during detection suggests a non-zero loss rate for the $\ket{r'}$ atom inside the ensemble. To obtain insight into the loss mechanism, we compare three different experimental sequences: (1) keeping both the probe and control light beams on for $\sim 30$~$\mu$s (green curve in Fig. S2), (2) turning on the control light $20$~$\mu$s earlier than the probe light (purple curve), and (3) turning both control and probe light on at the same time, but $20$~$\mu$s later than in case (1) (tan curve).

For sequences (1) and (3), the photon transmission levels at the time when the probe is turned on are approximately the same. This excludes the possibility that the loss of $\ket{r'}$ is due to the expansion of the gas or to other environment noise during the detection stage. On the other hand, there is a significant difference between the data for sequences (2) and (3). The larger loss for situation (2) indicates that the control light beam induces loss of $\ket{r'}$, possibly due to coupling of the Rydberg state to the unstable $5P_{3/2}$ state, as explained in the main text. However, there is some remaining difference between the transmission for sequences (1) and (2), which suggests a smaller additional loss during detection that requires both the control and the probe light to be applied. The extra loss mechanism may be related to the autoionization process that can occur when the Rydberg polariton with Rydberg excitation component in $\ket{r}$ collides with the Rydberg atom in $\ket{r'}$ \cite{robicheaux2005ionization, li2005dipole,kiffner2016quantum,mizoguchi2019ultrafast}.

We have verified that photoionization rate from the control beam is too small. According to Ref. \cite{saffman2005analysis}, the expected photoionization rate under our conditions is $\Gamma_{pi} \approx 340$~s$^{-1}$, two orders of magnitude smaller that the observed rate constant $\Gamma_{tr}=3.5 \times 10^4$~s$^{-1}$ for the transmission curve. Furthermore, the repulsive ponderomotive potential of the control beam is too small by three orders of magnitude to explain the increase in transmission with time. On the other hand, a dc electric field of 20-40~mV/cm could provide enough admixture of the 92S Rydberg state to the 91P state so that the state can be coupled with the control laser to $5P_{3/2}$, from where it decays to the ground state.

\begin{figure}[htbp]
\centering
\includegraphics[width=1\textwidth,trim= 0 200 150 50,clip]{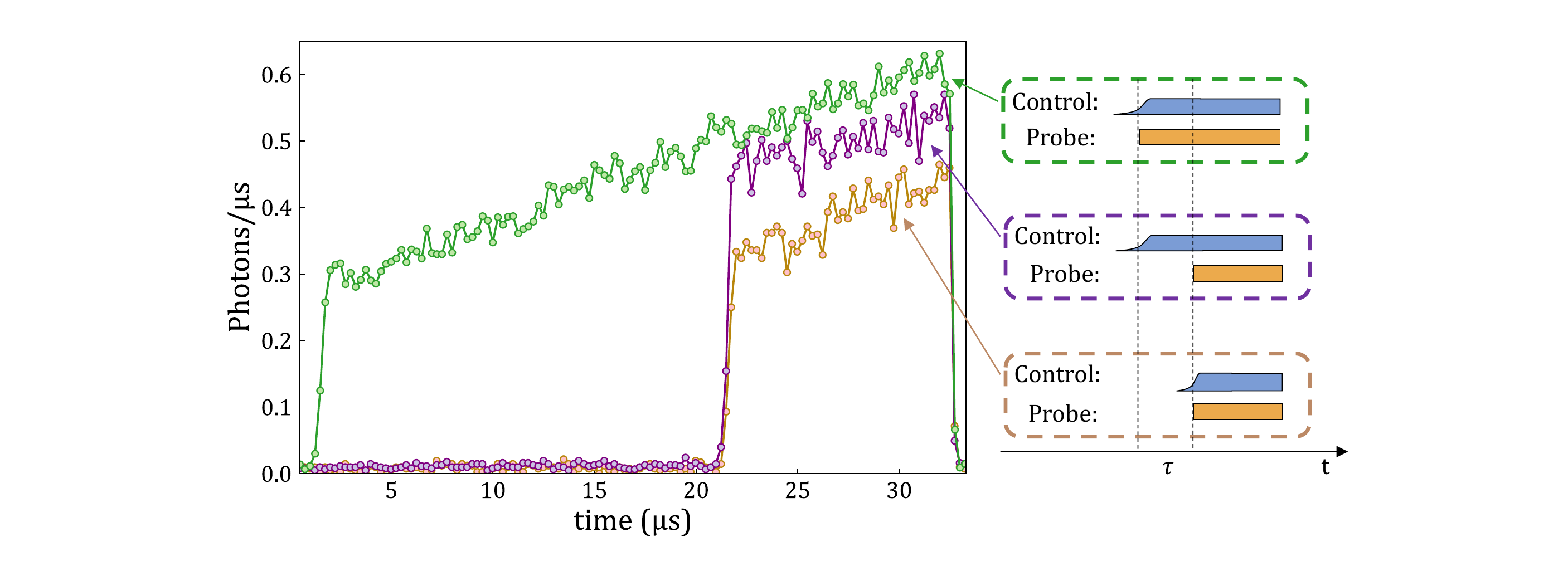}
\caption{\label{fig:Ramsey}
Three different experiment sequences, as illustrated by the schematic figures on the right, are used to investigate the mechanisms for the loss of $\ket{r'}$ during the detection stage. Green points represent the data where both control and probe light are turned on for 35~$\mu$s, purple points represent the data when control light was on for the entire 35~$\mu$s and probe light is turned on only for the last 15us. Tan points represent the data when both control and probe light are turned on only for the last 15~$\mu$s. From the photon rate level of each sequence we can conclude that control light has the most effect in increasing the loss of the $\ket{r'}$ state, and hence the transmission over time.  }
\end{figure}

\begin{figure}[htbp]
\centering
\includegraphics[width=1\textwidth ,trim= 0 100 0 100,clip]{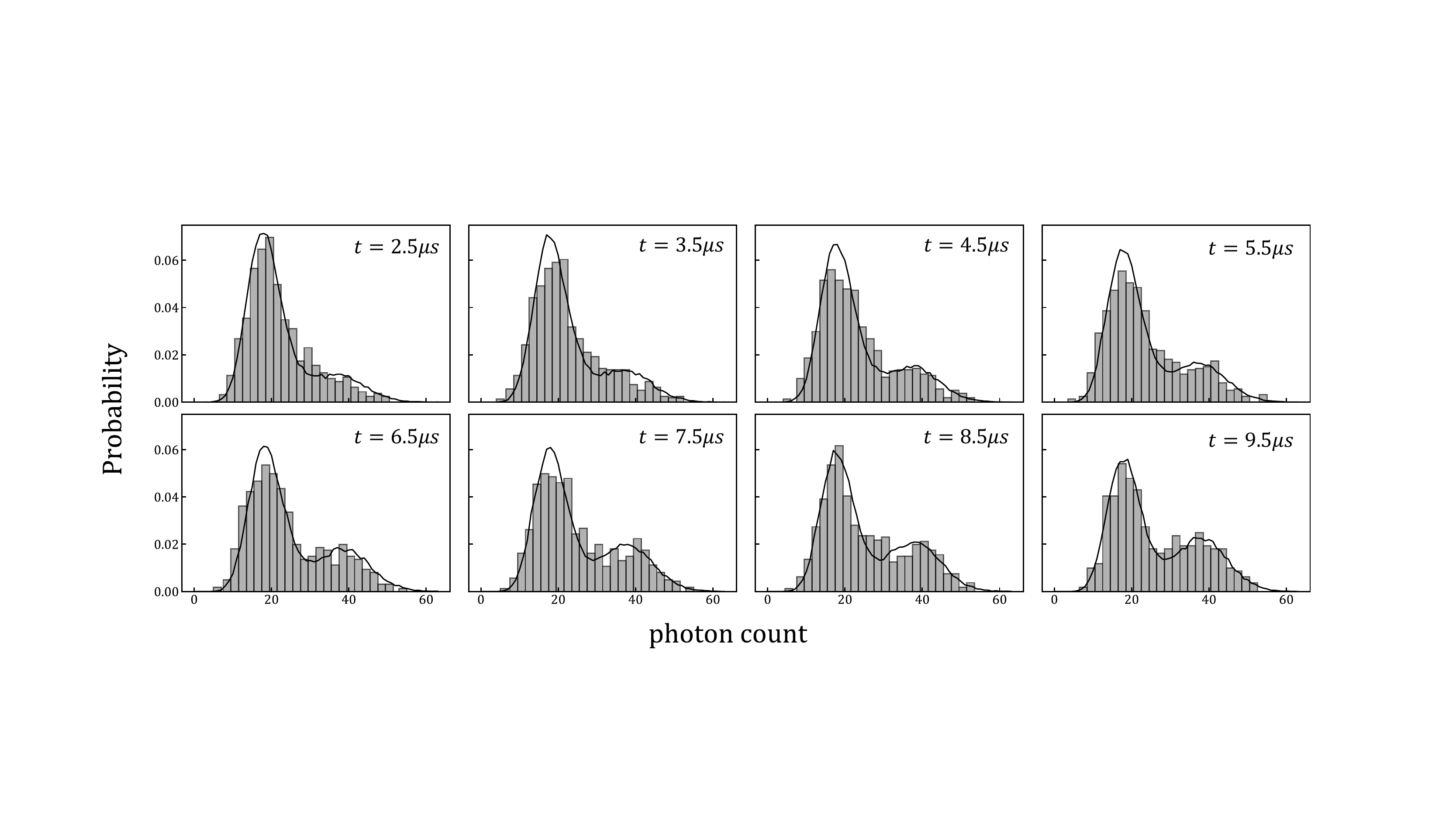}
\caption{\label{fig:prepFid}
Measured distributions at varying start times of probing window with an integration time of 6 $\mu s$. These distributions are fit using our theoretical model to obtain loss rate and preparation fidelity. The inferred preparation fidelity is $F_p=0.93\pm 0.02$
}
\end{figure}

We model the loss of the Rydberg atom in $\ket{r'}$ during detection as a sudden change in the transmission. We assume that this occurs randomly with constant probability. We then use this model to predict the expected histogram of probe photon counts at various detection times and compare it to the experimental data (Fig. \ref{fig:prepFid}). Our model has four free parameters: the photon rates in the presence and absence of an atom in $\ket{r'}$, respectively, the loss rate for the state $\ket{r'}$, and the fidelity of initially preparing the atom in $\ket{r'}$. We fit the experimental data for various detection times, as shown in Fig. \ref{fig:prepFid}, and find a loss rate constant of $\Gamma_{tr}=0.035$~$\mu$s$^{-1}$, and a preparation fidelity of $F_p=0.93 \pm 0.02$. The model also captures the average rise in transmission that we observe, as shown in Fig. 2(b) of the main text. 

We use a similar model to fit the photon histogram for no atom initially prepared in the Rydberg state, where we assume that a Rydberg impurity is created at random times due to the decay of the Rydberg polariton during detection. This process manifests itself as a sudden increase in probe transmission. We fit the histogram and the average transmission to find a Rydberg impurity creation rate of 0.015~$\mu$s$^{-1}$.

The creation of the Rydberg impurity, combined with the self-blockade effect, limits the maximum probe photon rate. While a higher photon rate is preferred for faster detection, the fidelity eventually reduces as the probe photon rate increases. Fig. S4 plots the detection fidelity at various photon rates. We choose a photon rate of $8$ $\mu$s$^{-1}$ for the measurements shown in the main text.

\begin{figure}[htbp]
\centering
\includegraphics[width=0.5\textwidth ,trim= 0 50 0 50,clip]{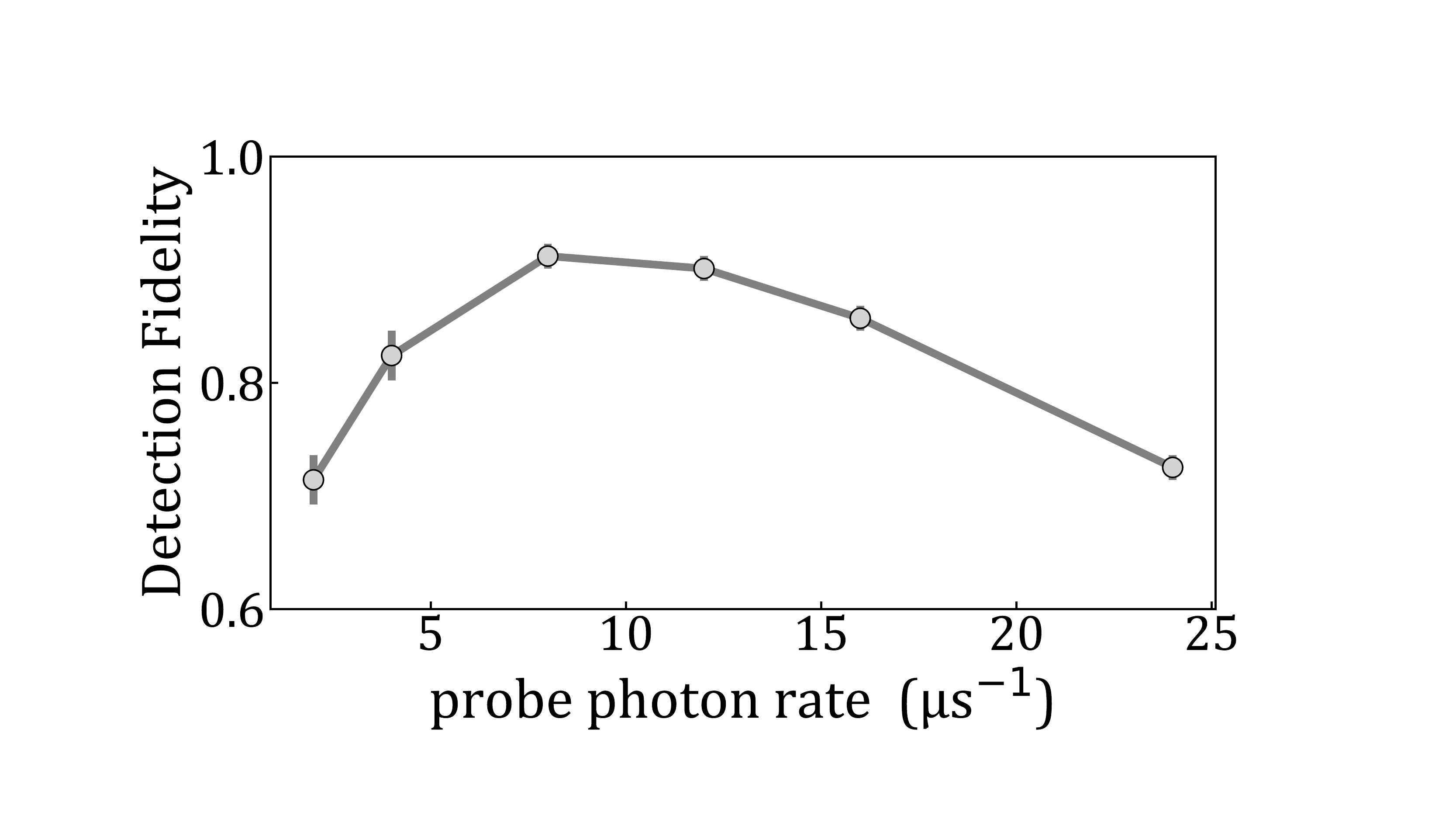}
\caption{\label{fig:photonRate}
Fidelity of detection as a function of the detected probe photon count rate
with optimized readout time for each data point (between 3~$\mu$s and 8~$\mu$s). The data have been corrected for the preparation fidelity.
}
\end{figure}

\subsection{Repeated (non-destructive) detection}
In the main text we provided our readout fidelity of $F_d=0.92 \pm 0.04$ in a $6$ $\mu$s readout window, and for $30$ counts as a detection threshold. If we perform subsequently a second measurement in $6$ $\mu$s with the same threshold, we find for this window a detection fidelity of $0.81 \pm 0.04$ ($\approx F_d^2$). Fig. 2c of the main text shows the correlations between the two measurements; the results are summarized in Table I. This correlation between consecutive measurements is a signature of quantum nondemolition measurements (QND) (\cite{QND_NV}, \cite{QND_review}), and under ideal conditions we would not expect the measurement to induce any change in the state. The average conditional probability that the second measurement yields the same result as the first measurement is $0.79 \pm 0.03$.

\begin{table}[ht]
    \centering
    \begin{tabular}{ |c|c|c|c|c|  }
        \hline
        Measurement & \multicolumn{2}{c|}{Prepare $\spinup$} & \multicolumn{2}{c|}{No $\spinup$ preparation} \\
        \cline{2-5}
         & Detect $\spinup$  & No $\spinup$ detection & Detect $\spinup$ & No $\spinup$ detection\\
        \hline
        first &0.92  & 0.08    & 0.10  & 0.90 \\
        \cline{1-5}
        second & 0.76 & 0.24 & 0.19 & 0.81\\
        \hline
    \end{tabular}
    \caption{Results for repeated (non-destructive) measurement. $\spinup$ preparation and detection refer to the preparation of an atom in the Rydberg state $\ket{91P_{3/2}, m_j=1/2}=\ket{r'}$. Both the first and second measurements last for 6 $\mu$s. The error (standard deviation) for each element in the table is 0.04.}
    \label{NonDestructiveReadout}
\end{table}

\subsection{Measurements of Rabi oscillations}
We use specific magnetic sublevels  $\ket{r'}=\ket{91P_{3/2},m_J=3/2}$ and $\ket{r}=\ket{92S_{1/2},m_J=1/2}$ to define our qubit. However, there is also the possibility of off-resonant coupling to other magnetic sublevels, especially at high Rabi frequency. We apply a magnetic field of $9$~G to lift the Zeeman degeneracy, which results in a Zeeman splitting between neighboring magnetic sublevels of $\sim 17$~MHz for the $\ket{91P_{3/2}}$ manifold, and $\sim 25~$MHz for the $\ket{92S_{1/2}}$ manifold. To reduce the coupling to other transitions by the the microwave field, we use two radio frequency antennas whose relative amplitude and phase are tuned to suppress the $\pi$-polarized transition $\ket{92S_{1/2},m_J=1/2} \leftrightarrow \ket{91P_{3/2},m_J=1/2}$. We observe a suppression by a factor of 10 compared to a single antenna. In the future, a third microwave antenna can be added to also eliminate the $\sigma^{-}$ transition.



The normalized population of $\ket{r'}$ shown in the Rabi oscillation measurements (Fig.3 in the main text) is corrected by the preparation infidelity ($1-F_p$) and the detection infidelity ($1-F_d$). The measured probability $\tilde{p}(r')$ of detecting $\ket{r'}$  is related to the actual probability $p(r')$ by the following relation: $\tilde{p}(r') = F_p[(1-F_d)+(2F_d-1)p(r')]$, which is used to remove the preparation and readout error. Moreover, during the 2-hour measurements of the Rabi oscillations, there were drifts in alignments, which affected the averaged transmitted probe photon number. To account for such long-time drift, we added two extra reference measurements: one without microwave driving field, and another one without preparation stage. These two measurements allow us to monitor any slow drift of $F_p$ and $F_d$, and re-scale the observed state probabilities.

\subsection{Rydberg blockade}

\textbf{Preparation blockade:}
During our preparation stage, we couple atoms to the $\ket{r'}\def \ket{91P_{3/2}, \text{m}_{J}=3/2}$ state. A pair of atoms in this state will experience an anisotropic van der Waals interaction $V_{r'r'}= C_{6}(\theta)/|R|^{6}$, where $R$ is the distance between the pair of atoms and $\theta$ is the angle the between the pair of atoms and the quantization axis (Fig. \ref{fig:PPvdw}a). From this interaction we can estimate the effective blockade radius $V_{r'r'}(r_{B}(\theta)) = \hbar\Gamma_{3}/2$ where $\Gamma_{3}$ is the full width half maximum of our 3-photon resonance, $\Gamma_{3}= 2 \pi \times 0.6$ MHz. We estimate a blockade radius of about $12~\mu$m along the quantization direction ($\theta = 0$). The resulting blockade volume has an ellipsoid shape with an aspect ratio of $1.6$, as shown on Fig. \ref{fig:PPvdw}b. The average blockade radius of the ellipsoid is $r_{B} =15~\mu$m.   

\begin{figure}[htbp]
\centering
\includegraphics[width=0.75\textwidth ]{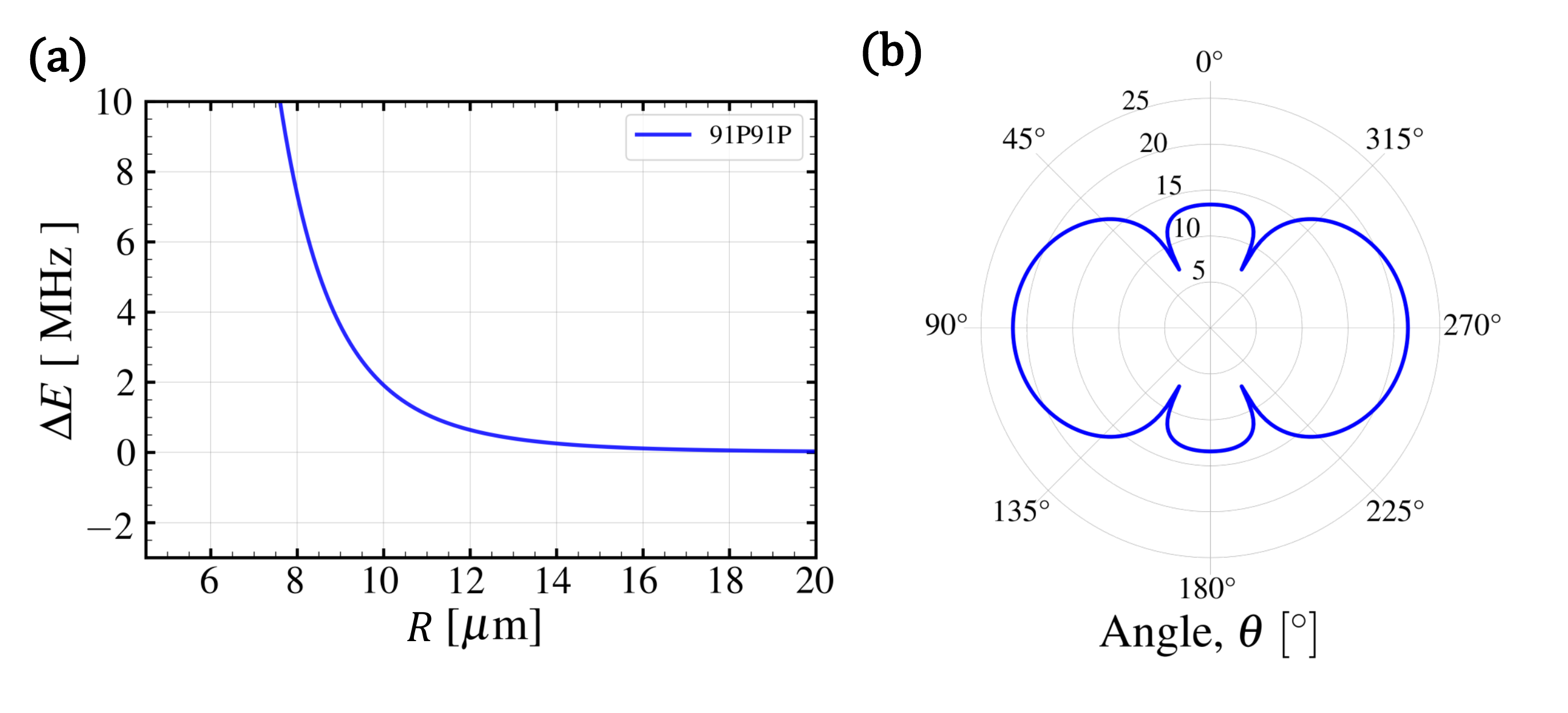}
\caption{\label{fig:PPvdw}
\textbf{Rydberg blockade during preparation:} (a) Interaction energy $\Delta E$ between two atoms in state $\ket{r'}$  as a function of separation $R$ at $\theta=0$ calculated from exact diagonalization of the interaction Hamiltonian. The fitted van der Waals interaction coefficient is $C_{6}\left(\theta=0\right) = 2\pi\times1.94~\text{THz} \cdot \mu\text{m}^6$, which results in a blockade radius of $r_{B}\left( \theta=0\right) = 12~\mu$m. (b) $r_{B}\left(\theta\right)$ vs. $\theta$. Due to the anisotropy of the atomic wave-functions involved, the resulting interaction resembles an ellipsoid with aspect ratio of $1.6$, with the semi-major axis at $90\degree$ from our quantization axis. This curve was obtained by perturbatively calculating $C_6(\theta)$.
}
\end{figure}

\textbf{Detection blockade:}
During the detection stage, we couple our ground state atoms to the $\ket{r}\def \ket{92S_{1/2}, \text{m}_{J}=1/2}$ state. Since $r$ and $r'$ have different parity they are dominated at large distances by dipole-dipole interactions $(\sim R^{-3})$, while at small distances are dominated by van der Waals interactions $(\sim R^{-6})$. The dipole-dipole interaction induces the formation of symmetric and anti-symmetric molecular states ($\ket{\pm}=\ket{rr'}\pm\ket{r'r} / \sqrt{2}$). The blockade radius during the detection stage is defined as $V_{\pm}(r_{B\pm})=\hbar\Gamma_{EIT}/2$, with $\Gamma_{EIT}$ being the EIT linewidth. The estimated blockade radius for each branch are $r_{B+}=12.7~\mu$m and $r_{B-}=6.2~\mu$m respectively. The estimated blockaded radius is therefore the average radius from both branches $r_{B} = 9.4~\mu$m. 

\begin{figure}[htbp]
\centering
\includegraphics[width=0.6\textwidth ]{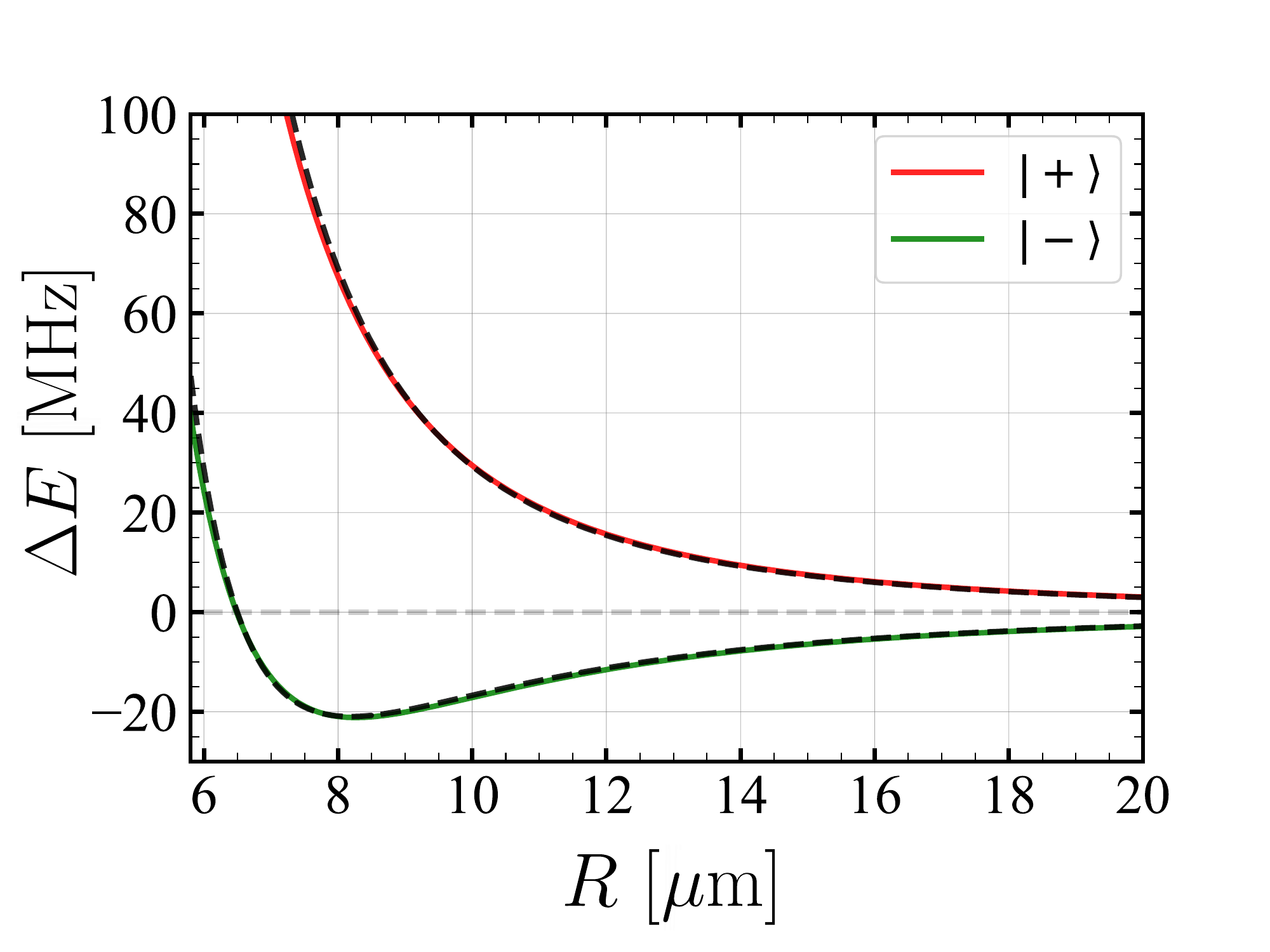}
\caption{\label{fig:SPvdw}
\textbf{Rydberg blockade for $\ket{rr'}$ during detection:} (a) $\ket{r r'}$ pair state interaction energy $\Delta E$ vs. separation $R$ at $\theta=0$ calculated from exact diagonalization of the interaction Hamiltonian. The dipole-dipole interaction induces the formation of symmetric (red) and anti-symmetric (green) states $\ket{\pm}=(\ket{rr'}\pm\ket{r'r} )/ \sqrt{2}$. Black dashed lines correspond to the fitted model $A/R^{6}\pm 
B /R^{3}$, where the fit gives $A=C_{6}/h=6310~\text{GHz} \cdot \mu\text{m}^{6}$ and $B=C_{3}/h=23.6~\text{GHz}\cdot \mu\text{m}^{3}$.
}
\end{figure}

\bibliographystyle{apsrev4-1}
\bibliography{reference}